\documentclass[prl,twocolumn,showpacs,amsmath,amssymb]{revtex4}

\usepackage{amsmath}
\usepackage{graphicx}
\usepackage[pdftex,dvips]{epsfig}
\newcommand{\be}{\begin{equation}}
\newcommand{\ee}{\end{equation}}
\newcommand{\bma}{\begin{displaymath}}
\newcommand{\ema}{\end{displaymath}}
 
\begin{document}

\title{Edge-dependent selection rules in magic triangular graphene flakes}

\author{J. Akola, H.P. Heiskanen, and M. Manninen}

\affiliation{\sl NanoScience Center, Department of Physics, P.O. Box 35
FIN-40014 University of Jyv\"askyl\"a, Finland}

\date{\today}

\begin{abstract} 

The electronic shell and supershell structure of triangular graphene quantum 
dots has been studied using density functional and tight-binding methods. The 
density functional calculations demonstrate that the electronic structure close 
to the Fermi energy is correctly described with a simple tight-binding model 
where only the $p_z$ orbitals perpendicular to the graphene layer are included. 
The results show that (i) both at the bottom and at the top of the $p_z$ band a 
supershell structure similar to that of free electrons confined in a triangular 
cavity is seen, (ii) close to the Fermi level the shell structure is that of free 
{\it massless} particles, (iii) triangles with armchair edges show an additional 
sequence of levels ('ghost states') absent for triangles with zigzag edges while 
the latter exhibit edge states, and (iv) the observed shell structure is rather 
insensitive to the edge roughness.

\end{abstract}

\maketitle

Recent experimental success in manufacturing single layer graphene flakes 
on various surfaces \cite{berger2004,novoselov2004,novoselov2005,berger2006} 
has made graphene a new playground for theoretical and computational 
physics, \cite{alicea2005,gusynin2005,gusynin2006,nomura2007,son2007} and 
more and more experimental results are emerging. 
\cite{novoselov2007,geim2007,li2007} Most of the recent interest has been
focused in the effects caused by the peculiar band structure of graphite 
near the Fermi level ($\epsilon_{\rm F}$): Electrons and holes behave as 
massless particles (Dirac fermions) due to the linear dispersion relation 
although their velocity is very small. \cite{zhou2006}

The triangular shape of two-dimensional clusters is particularly interesting 
because, in the case of free electrons, it supports perhaps the most persistent 
and regular supershell structure of all systems.\cite{brack1997} Furthermore, 
the triangular shape is preferred in two-dimensional metallic systems
\cite{kolehmainen1997,jenssens2003} and in plasma clusters.\cite{reimann1998} 
For tetravalent elements, triangular clusters have been observed in silicon. 
\cite{lai1998} It is reasonable to expect that such shapes can be observed 
also for carbon, and this is supported further by the fact that equilateral 
triangles of graphene can be cut with the two most stable edge structures, the 
zigzag edge and the armchair edge. 

In this letter, we wish to point out that finite graphene flakes (or quantum 
dots) have an intriguing energy spectrum close to the Fermi level. We have
performed electronic structure calculations for triangular graphene flakes using
the density functional theory (DFT) for all the valence electrons, and a 
tight-binding (TB) approach that considers only the carbon $p_z$ electrons 
(H\"uckel model). Our results show that already in small triangular flakes 
($N$=300, $L$=$5$ nm) the electronic levels close to $\epsilon_{\rm F}$ can be 
understood as those of free massless electrons confined in a triangular cavity. 
Especially, we demonstrate that the edge structure has a selective role in the 
electronic shell structure: The zigzag edge prohibits a whole sequence of 
localized states {\it inside} the cluster although it supports edge states. This 
leads to well-defined edge-dependent selection rules that are based on an 
analytical model. Recently, Yamamoto {\it et al.} addressed the presence 
(absence) of edge states at $\epsilon_{\rm F}$ in zigzag (armchair) triangles 
of graphene, and the effect on the optical absorption, \cite{yamamoto2006} but 
the simple principles of the underlying energy spectrum have remained 
unexplained.

\begin{figure}[h]
\vskip -1cm
\includegraphics[angle=-90,width=0.8\columnwidth]{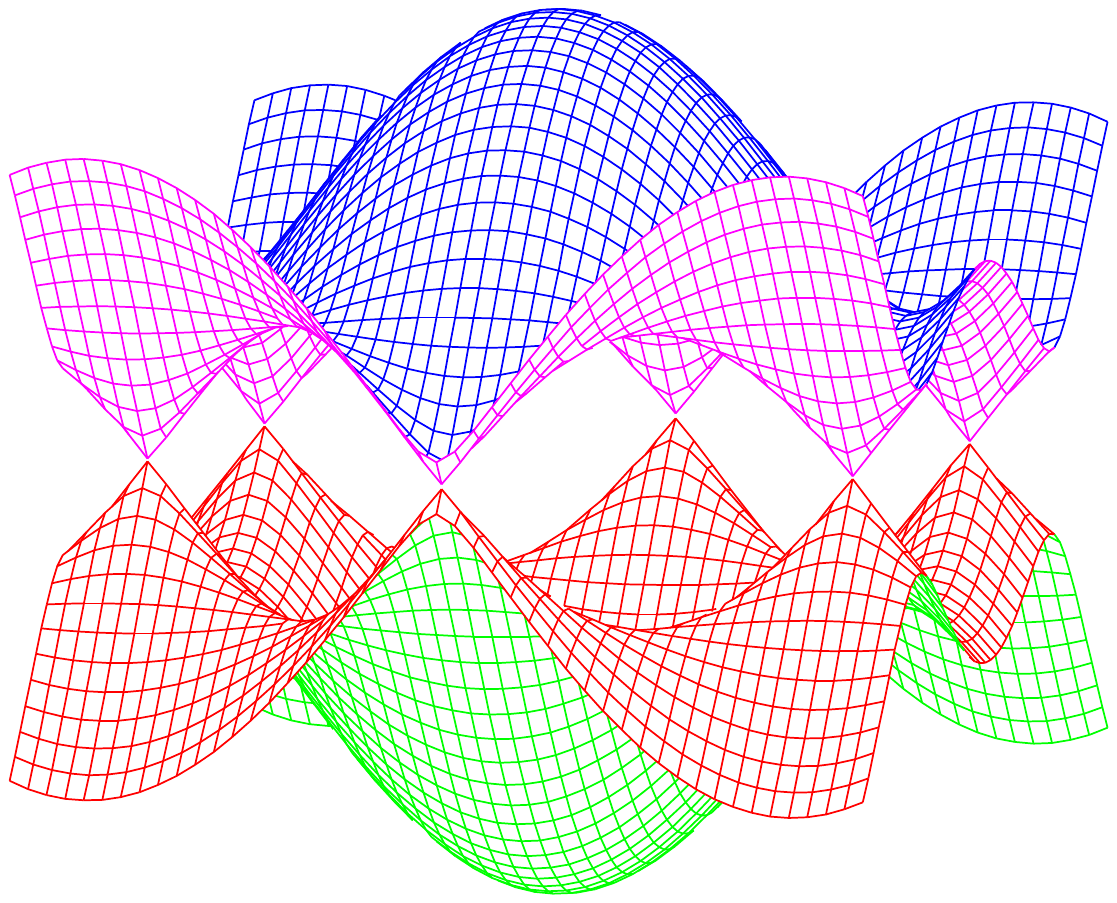} 
\vskip -2cm
\includegraphics[width=0.8\columnwidth]{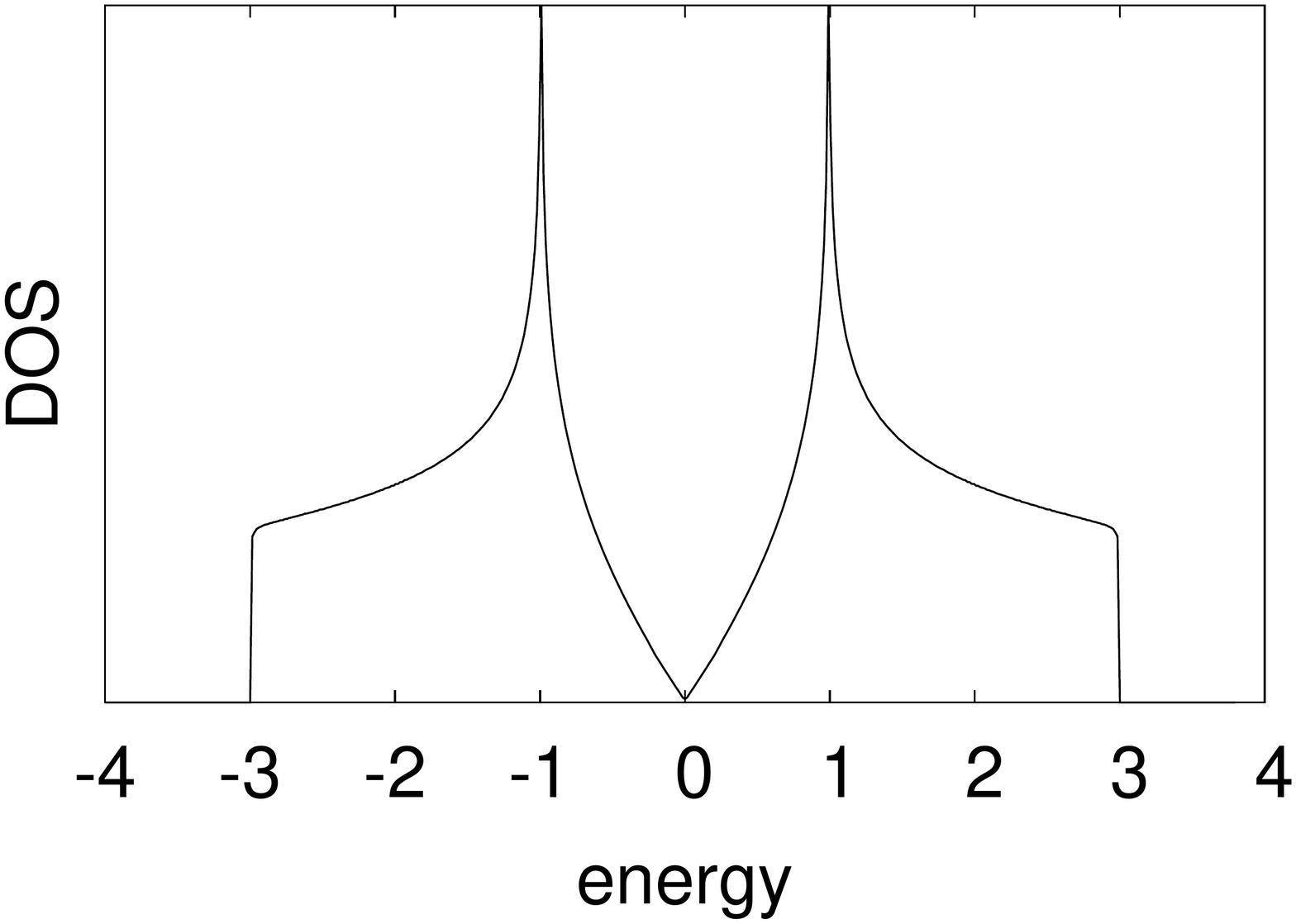}
\caption{Cross-over of the valence and conduction bands at the Fermi 
energy (top), and the density of states (bottom, $p_z$ electrons) of an 
infinite graphene sheet.}
\label{bands}
\end{figure}

It is well-known that the atomic $p_z$ electrons perpendicular to the 
graphene plane are responsible for the captivating band structure shown in 
Fig. \ref{bands} with the valence the conduction bands meeting at the 
corners of the hexagonal Brillouin zone. \cite{elliot1998,areshkin2007} The 
Fermi surface consists of a discrete set of these points of high-$k$ value, 
and the resulting density of states (DOS) has a zero weight at 
$\epsilon_{\rm F}$. The cross-over regions have locally hourglass-like shapes, 
which results in the linear and isotropic electron dispersion relation in the 
conduction band ($\epsilon>\epsilon_F$=0) but only in a small energy interval. 
Since the atomic $p_z$ electrons are perpendicular to the graphene plane 
their interaction with the neighboring atoms does not have any directional 
dependence and, consequently, they can be described as $s$-type electrons in 
the TB model. Neglecting also the differential overlap between atomic sites 
the system can be described with the traditional H\"uckel model
\bma
H_{ij}=\left\{ \begin{array}{rl}
-t, & {\rm if} \quad i,j \quad{\rm nearest \quad neighbours}\\
0, & {\rm otherwise}\\
\end{array}\right. ,
\ema
where the hopping parameter $t$ (resonance integral) determines the width 
of the bands and the on-site energy is chosen to be $\epsilon_F=0$. We choose 
to present our results in units $t=1$. The resulting TB bands (Fig. \ref{bands}) 
reaches from -3 to +3 (in real graphene our unit $t$ corresponds to about 2.6 eV).

A conceptual cutting of a finite graphene flake breaks covalent bonds yielding 
edges with dangling bonds. We consider the dangling bonds to be passivated, say 
with hydrogen. Since the covalent bonding with hydrogen involves $sp^2$ 
hybridized orbitals, the passivation is expected to have only a small effect on 
the perpendicular $p_z$ electron states. Therefore, we neglect this effect in 
our TB model and follow Areshkin {\it et al.} \cite{areshkin2007} and treat the 
edge atoms in the same footing as bulk atoms. Moreover, we will completely 
neglect the interaction of graphene with the possible substrate and treat the 
graphene flake as an isolated two-dimensional cluster or quantum dot. As we 
shall see, the results of the simple TB model agree well with those of the full 
DFT calculation.

\vskip0.3truecm

\begin{figure}[h]
\vskip -3cm
\includegraphics[width=\columnwidth]{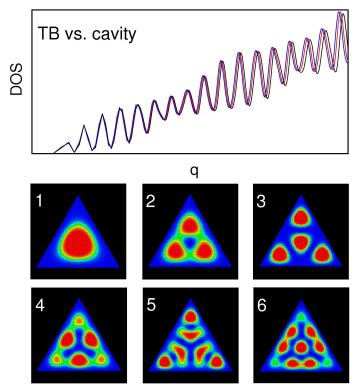}
\vskip -1cm
\caption{Upper panel: DOS at the bottom of the TB band shown as a function of  
$q=\sqrt{\epsilon+3t}$. Blue: zigzag triangle with 10000 atoms; red: armchair 
triangle with 9918 atoms; and black: result for free electrons in a triangular
cavity. Lower panel: Electron densities of the six lowest energy levels.}
\label{dos1}
\end{figure}

It has been shown that at the bottom of the valence band the TB model gives 
exactly the free electron states for a triangular lattice,\cite{manninen1991} 
and the same is true also for the hexagonal graphene. Consequently, at the 
bottom (and at the top) the energy levels are expected to show the same shell 
structure as free electrons in a triangular cavity, determined by the 
equation \cite{borghis1958,krishnamurthy1982}
\be
\epsilon_{n,m}=\epsilon_0(n^2+m^2-nm),
\label{levels}
\ee
where $\epsilon_0=8\pi^2\hbar^2/3m_eL^2$, $L$ being the length of the triangle 
side. The quantum numbers must satisfy $m\ge 1$ and $n\ge 2m$. Determination
of the electron effective mass in the graphene lattice gives for the TB model 
$\epsilon_0=4\pi^2t/9N$, where $N$ is the number of atoms in the triangle 
($L=3d\sqrt{N}/2$ for a large triangle, $d$ is the nearest-neighbor distance).

The shell structure manifests itself as a regular variation of 
the density of the states (DOS) which can be determined by Gaussian
convolution of the discrete levels.
Figure \ref{dos1} shows DOS close to the bottom of the valence band obtained 
from the above equation and compared to the TB model for two graphene
triangles, one with 10000 atoms (zigzag edge) and the other with 9918 atoms 
(armchair edge). The profiles are clearly similar and exhibit the beating 
pattern of the supershell structure\cite{reimann1998}. Note that DOS is plotted 
as a function of $\sqrt{\epsilon+3t}$ making the shells equidistant. Fig. 
\ref{dos1} shows also the electron densities corresponding to the six lowest 
energy levels (for degenerate states we show the sum of the density). The 
density patterns are identical to those of free electrons confined in a 
triangle \cite{jenssens2003} or wave modes in triangular resonators 
\cite{huang2001}.

The Fermi level of graphene consists of two equivalent points at the border 
of the Brillouin zone (see Fig. \ref{bands}) where the conduction and valence
bands open as circular cones resulting a linear dispersion relation for 
electrons $\epsilon({\bf k})=C\hbar k$, where $C$ is the velocity. Thus, it is to be 
expected that the electron dynamics is not determined by the Schr\"odinger 
equation but by the wave equation of massless particles (or the Dirac equation). 
For free particles confined in a triangle the energy levels are still 
determined by Eq. (\ref{levels}) but now it results in the square of the energy, 
i.e
\be
\epsilon_{n,m}=\epsilon_1\sqrt{n^2+m^2-nm},
\label{levels2}
\ee
where $\epsilon_1=2\pi t/\sqrt{3N}$. It is interesting to note that these energy 
levels were actually computed for the wave equation much 
earlier than for the Schr\"odinger equation. \cite{borghis1958}

\begin{figure}[h]
\vskip-3cm
\includegraphics[width=\columnwidth]{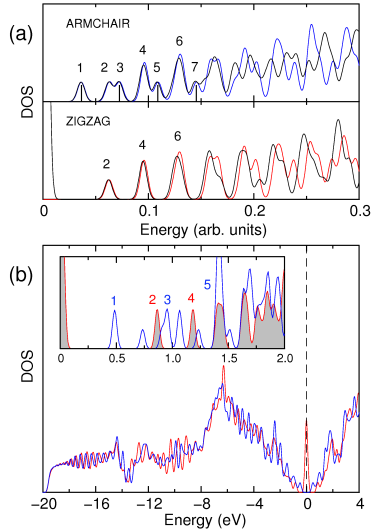}
\caption{Upper panel: TB-DOS at the Fermi level displayed as a function of 
energy (red and blue curves) compared to the density of levels of Eq. 
(\ref{levels2}) (black curves). The zigzag triangle has 10000 atoms and the 
armchair triangle 9918 atoms. Lower panel: DOS of the full DFT calculation 
for the triangular C$_{321}$H$_{51}$ (zigzag, red) and C$_{330}$H$_{60}$ 
(armchair, blue) flakes. The inset shows the levels above the Fermi surface
where the zigzag spectrum is shaded.}
\label{dos2}
\end{figure}

Figure \ref{dos2} shows TB-DOS above the Fermi energy for two large triangles
($\sim 10000$ atoms) with zigzag and armchair edges and compares them with 
the levels of free massless electrons [Eq. (\ref{levels2})]. The results are
the following: (i) Each energy level has an additional degeneracy of two due 
to the two equivalent points at $\epsilon_F$. (ii) The zigzag triangle shows 
the levels of Eq. (\ref{levels2}) with index values $m\ge 1$ and $n\ge 2m$ 
while the armchair edge shows all the levels where $n\ge m \ge 1$. (iii) The 
states are much less dense than at the bottom of the band, and Eq. 
(\ref{levels2}) describes only the lowest states accurately. (iv) Due to the 
sparseness of the states, no supershell oscillations are visible for the 
massless particles (although the supershell structure of ordinary electrons 
is clearly seen in Fig. \ref{dos1}). (v) The zigzag edge supports particularly 
visible edge states \cite{kobayashi1993,nakada1996} that appear at $\epsilon_F$ 
as a prominent peak. The number of these states equals the number of the 
outermost edge atoms in zigzag triangles, which is $N_{\rm ss}=\sqrt{N}$. 

In order to compare our results with a more realistic calculation, we have 
performed DFT calculations for triangular C$_{321}$H$_{51}$ (zigzag) and 
C$_{330}$H$_{60}$ (armchair) flakes with the CPMD program.\cite{CPMD} The 
DFT calculations use a plane wave basis set ($E_{cut}=50$ Ry), pseudopotentials,
\cite{TM91} and a generalized gradient-corrected PBE approximation for the 
exchange-correlation energy. \cite{PBE96} The resulting DFT-DOS of {\it all} 
valence electrons is plotted in Fig. \ref{dos2}(b) for both systems, and they 
show overall features characteristic for graphite. The zigzag edge states at 
$\epsilon_F$ are visible, and the closest conduction states obey the simple 
analytical model of Eq. (\ref{levels2}). The even-numbered peaks are split for 
the armchair triangle, which is a result reproduced by TB (the splitting 
reduces with increasing system size).

\begin{figure}[h]
\vskip -3cm
\includegraphics[width=\columnwidth]{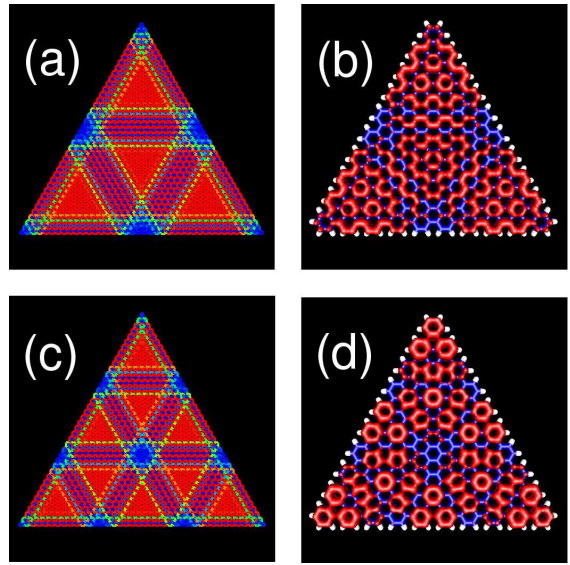}
\caption{Electron density of the 3rd [(a) and (b)] and 5th [(c) and (d)] 
energy levels above the Fermi energy in armchair triangles ('ghost states', 
labeled in Fig. \ref{dos2}, each has a degeneracy two). (a) and (c) are 
computed for a large TB triangle of 4920 C atoms while (b) and (d) are DFT 
results for a C$_{330}$H$_{60}$ molecule.}
\label{ghost}
\end{figure}

\begin{figure}[h]
\vskip -3cm
\includegraphics[width=\columnwidth]{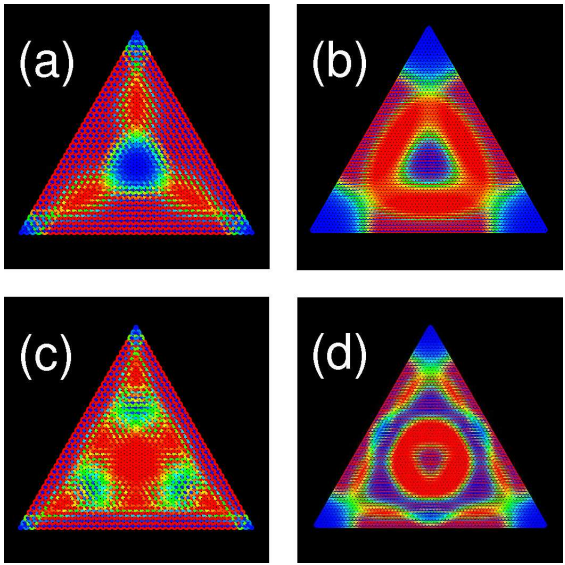}
\caption{Electron density (TB model) of the 2nd [(a) and (b)] and 4th [(c) 
and (d)] energy levels above the Fermi energy (labeled in Fig. \ref{dos2}) 
for armchair and zigzag triangles of 4920 and 5181 C atoms, respectively.}
\label{dens}
\end{figure}

The lowest conduction states that are labeled in Fig. \ref{dos2} show 
fascinating details, and the electron densities of two such states are 
visualized in Figure \ref{ghost}. For comparison, we show the same 
states/orbitals calculated for a large triangle with the TB model (4920 C 
atoms) and for a small triangle calculated with the DFT method (330 C atoms). 
The internal structure (symmetry) of the states is clearly similar, and 
therefore, it is independent of the triangle size and the model used. The 
states close to the Fermi level appear very different from those at the bottom 
of the band (Fig. \ref{dos1}). They are {\it not} simple densities of massless 
particles confined in a triangle since the density profile does not decay to 
zero at the edges. The corresponding electron levels are close to the 
Brillouin zone boundary, having large $k$-values, and the wave 
functions have pronounced oscillations with wave lengths that are related to 
the unit cell size. These oscillations guarantee that the wave function 
will be formally zero at the edges, but the corresponding pseudowave 
function of the massless particle does not necessarily show the same behavior. 
An interesting feature in Fig. \ref{ghost} is that the states have simple 
geometric structure of triangular symmetry. The size (number) of the triangles 
decreases (increases) with increasing energy, i.e. the pattern repeats itself. 
These 'ghost states' are completely absent for the zigzag triangles, and they 
correspond to quantum numbers of Eq. (\ref{levels2}) not allowed for free 
electrons in a triangular box [i.e. $2m\ge n\ge m\ge 1$ in Eq. (\ref{levels2})].

Figure \ref{dens} shows the electron densities corresponding to the 'normal' 
low energy states that obey the standard selection rules ($m\ge 1$ and $n\ge 2m$). 
Again, the electron density does not necessarily vanish at the edges of the 
triangle. The corresponding states for the armchair and zigzag triangles 
display obvious differences despite the fact that they involve the same set 
of quantum numbers (and energy). 

Finally, we want to note that a small roughness of the edge does not remove 
the peculiar states shown in Fig. \ref{ghost} or change the shell structure 
close to the Fermi level. These 'ghost states' form a triangular network, and 
it would be interesting to study if they can exist also in the graphene flakes 
with hexagonal, parallelogram, or trapezoidal shapes. 

In conclusion, we have computed the electronic structure of triangular graphene 
flakes and shown that the DOS profile close to $\epsilon_F$ is independent of the 
triangle size, and it can be described with the simple TB model. The zigzag flakes 
exhibit well-known edge states, and the armchair triangles show an additional set 
of 'ghost states' (different selection rules) where the corresponding electron 
density makes a triangular lattice. 
In large triangles of 5000-10000 C atoms, the energy levels can be 
described accurately by considering free massless particles confined in a triangular 
cavity. Presumably, the electronic states near the Fermi surface are not sensitive 
to the dielectric substrate, and we expect that these fascinating wave functions can 
be observed with scanning tunneling microscopy.

\end{document}